\documentclass{article}
\usepackage[T1]{fontenc} % add special characters (e.g., umlaute)
\usepackage[utf8]{inputenc} % set utf-8 as default input encoding
\usepackage{ismir}
\usepackage[export]{adjustbox}
\usepackage{amsmath}
\usepackage{amsfonts}
\usepackage{amssymb}
\usepackage{amsthm}
\usepackage{mathabx}
\usepackage{mathtools}
\usepackage{stmaryrd}

\usepackage{upgreek}

\usepackage{paralist}

\usepackage{cite}
\usepackage{url}

\usepackage{graphicx}

\usepackage{color}
\usepackage[table]{xcolor}

\usepackage{siunitx}
\usepackage{tabularx}
\usepackage{booktabs}

\newif\ifdoubleblind
\usepackage{lineno}
\usepackage{marginnote}

\newcommand{\ackedalex}[1]{}

\usepackage{collcell}
\usepackage{colortbl}
\usepackage{tikz}
\usepackage{lipsum}

\usepackage{xspace}

\usepackage[firstpage]{draftwatermark}
\definecolor{lightgray}{rgb}{0.9,0.9,0.9}
\definecolor{darkgray}{rgb}{0.4,0.4,0.4}
\SetWatermarkFontSize{12pt}
\SetWatermarkScale{1.1}
\SetWatermarkAngle{90}
\SetWatermarkHorCenter{202mm}
\SetWatermarkVerCenter{170mm}
\SetWatermarkColor{darkgray}
\SetWatermarkText{Late-Breaking / Demo Session Extended Abstract, ISMIR 2024 Conference}

\def\lbd{}	% Flag to use correct LBD settings in the paper, please do not modify this line

\title{Facing the Music: Tackling Singing Voice Separation\\in Cinematic Audio Source Separation}
\multauthor
{Karn N. Watcharasupat$^{1^*,2}$ \hspace{1cm} Chih-Wei Wu$^1$ \hspace{1cm} Iroro Orife$^1$} { 
$^1$ Audio Algorithms, Netflix Inc., Los Gatos, CA 95032, USA ($^*$ Internship)\\
$^2$ Music Informatics Group, Georgia Institute of Technology, Atlanta, GA 30332, USA\\
{\tt\small kwatcharasupat@gatech.edu, \{chihweiw, iorife\}@netflix.com}
}

\def\authorname{K.\ N.\ Watcharasupat, C.-W. Wu, and I. Orife}

\usepackage[bookmarks=false,%
    pdfauthor={\authorname},%
    pdfsubject={\papersubject},%
    hidelinks]{hyperref}

\usepackage{cleveref}
\crefname{figure}{fig.\!}{figs.\!}
\Crefname{figure}{Fig.\!}{Figs.\!}

\begin{document}

\maketitle

\begin{abstract}
Cinematic audio source separation (CASS), as a standalone problem of extracting individual stems from their mixture, is a fairly new subtask of audio source separation. A typical setup of CASS is a three-stem problem, with the aim of separating the mixture into the dialogue (DX), music (MX), and effects (FX) stems. Given the creative nature of cinematic sound production, however, several edge cases exist; some sound sources do not fit neatly in any of these three stems, necessitating the use of additional auxiliary stems in production. One very common edge case is the singing voice in film audio, which may belong in either the DX or MX or neither, depending heavily on the cinematic context. In this work, we demonstrate a very straightforward extension of the dedicated-decoder Bandit and query-based single-decoder Banquet models to a four-stem problem, treating non-musical dialogue, instrumental music, singing voice, and effects as separate stems. Interestingly, the query-based Banquet model outperformed the dedicated-decoder Bandit model. We hypothesized that this is due to a 
better feature alignment at the bottleneck as enforced by the band-agnostic FiLM layer.
\end{abstract}
\section{Introduction}\label{sec:introduction}

Applications of digital signal processing and machine learning to improve the audio experience in television or cinematic content have been attempted since at least the mid-2000s, often focusing on enhancing dialogue intelligibility with \cite{Paulus2019SourceSeparationEnabling} or without \cite{Uhle2008SpeechEnhancementMovie, Geiger2015DialogueEnhancementStereo} source separation as a preprocessing step. 
However, cinematic audio source separation (CASS), as a standalone problem of extracting individual stems from their mixture, remains an emerging subtask of audio source separation. To the best of our knowledge, most standalone CASS works thus far have relied on the three-stem setup introduced in \cite{Petermann2022CocktailForkProblem}, with the goal of separating the mixture into the dialogue (DX), music (MX), and effects (FX) stem. Although this is already a very useful setup for many downstream tasks, certain cinematic audio production workflow requires more granular controls over the distribution of sound sources into stems, with a significant number of edge cases and contextual nuances that evade putting all cinematic sound sources into three clear-cut boxes. As a result, additional auxiliary stems\footnote{See \href{https://tinyurl.com/nflx-mne-guidelines}{tinyurl.com/nflx-mne-guidelines}. Last accessed: 2 Aug 2024.} are often needed in some content, to account for sound events such as walla and singing voice.

Adding supports for these additional stems remains an open problem in CASS and poses a significant complexity increase in a system with a dedicated decoder for each stem. In this work,\footnote{Dataset and model implementation will be made available at \href{https://github.com/kwatcharasupat/source-separation-landing}{github.com/kwatcharasupat/source-separation-landing}.} we focus on adding support for distinguishing singing voice from speech and instrumental music, as this is perhaps one of the most unaddressed open problems in CASS research, as evidenced by the discussions in the Cinematic Demixing track of the 2023 Sound Demixing Challenge  \cite{Uhlich2023SoundDemixingChallenge}. To do so, we straightforwardly added an additional stem to the Bandit model \cite{Watcharasupat2023GeneralizedBandsplitNeural} and the Banquet model \cite{Watcharasupat2024StemAgnosticSingleDecoderSystem} and trained the systems on a modified version of the Divide and Remaster (DnR) v3 dataset \cite{Watcharasupat2024RemasteringDivideRemaster} with the music stem drawn from music source separation datasets to provide clean vocal and instrumental ground truths. Interestingly, the results indicated that the query-based single-decoder Banquet model consistently outperformed the dedicated-decoder Bandit model.

\section{Data}

Most CASS works so far relied on the DnR v2 \cite{Petermann2022CocktailForkProblem} dataset, which is a three-stem English-language dataset with the music stem drawn from the Free Music Archive (FMA) dataset \cite{Defferrard2017FMADatasetMusic}. The recently released multilingual rework (v3) of DnR \cite{Watcharasupat2024RemasteringDivideRemaster} also drew the music stem from FMA. Since FMA does not provide clean isolated vocal and instrumental stems, it is not possible to cleanly obtain isolated vocal and instrumental ground truths; there is also no way of distinguishing between types of vocalization (e.g. speech vs singing) in FMA. As a result, the dataset used in this work is an adaptation of DnR v3, with the music stems drawing from the music source separation datasets \mbox{MUSDB18-HQ}~\cite{Rafii2019MUSDB18HQUncompressedVersion} and MoisesDB \cite{Pereira2023MoisesDBDatasetSource} instead. We also removed all tracks and/or stems with bleed from these datasets. The generation process is similar to that in \cite{Watcharasupat2024RemasteringDivideRemaster}, with the vocals and instrumentals temporally aligned.

\section{System}

\begin{figure}
    \centering
    \hfill
    \includegraphics[width=0.4\linewidth,valign=t]{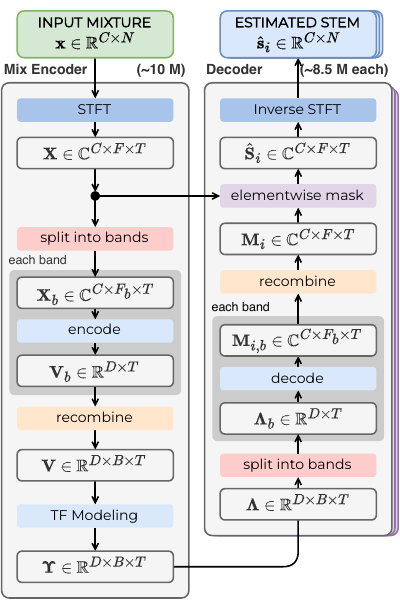}
    \hfill
    \includegraphics[width=0.4\linewidth,valign=t]{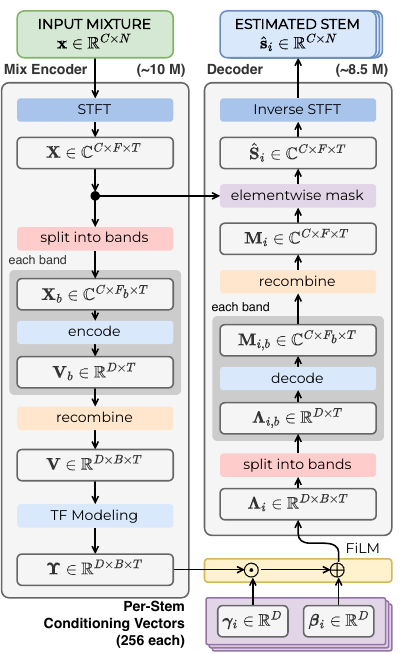}
    \hfill
    \caption{Model architecture for (left) Bandit and (right) Banquet. Bandit has a dedicated decoder for each stem. Banquet uses only one shared decoder.}
    \label{fig:architecture}
\end{figure}

The systems used in this work are (1) the 64-band musical Bandit model \cite{Watcharasupat2023GeneralizedBandsplitNeural} and (2) its query-based adaptation, Banquet \cite{Watcharasupat2024StemAgnosticSingleDecoderSystem}. The model architectures are shown in \Cref{fig:architecture}.  We experimented with three training setups. In the instrumental-only setup, the mixture does not contain any singing vocals. Each system extracts three stems: DX, \mbox{MX-I}, and FX. In the combined MX and split MX setups, the mixture contains singing vocals. The combined setup extracts three stems, with the MX stem containing both vocals and instrumentals (MX-*). The split setup extracts four stems, with singing vocals (MX-V) and instrumentals (MX-I) separately. Unlike the query-by-audio Banquet in \cite{Watcharasupat2024StemAgnosticSingleDecoderSystem}, the variant of Banquet used in this work is set up so that the conditioning vectors are directly learnable as stem-specific parameters. The training setup follows \cite{Watcharasupat2024RemasteringDivideRemaster}.

% Each model was trained for 200 epochs with a batch size of 8 per GPU. Each epoch consisted of 16384 training batches (2048 batches per GPU). For each training clip, a random chunk of \SI{8.0}{\second} was drawn for each stem independently of other stems. The training mixture chunks were then recomputed from the random stem chunks. Testing was done on the entire track, using overlap-add in the same way as in \cite{Watcharasupat2023GeneralizedBandsplitNeural, Watcharasupat2024StemAgnosticSingleDecoderSystem, Watcharasupat2024RemasteringDivideRemaster}, with a chunk size of \SI{8.0}{\second} and a hop size of \SI{1.0}{\second}. An AdamW optimizer \cite{Loshchilov2018DecoupledWeightDecay} with an initial learning rate of \num{e-3} and a decay factor of \num{0.99} per epoch was used. Training was done using 8 NVIDIA A100 Tensor Core GPUs (40 GB each) on an Amazon EC2 {p4d.24xlarge} instance. Testing was done using NVIDIA T4 Tensor Core GPUs on an Amazon EC2 {g4dn.metal} instance. 

\section{Results}

To evaluate the performance of the models, we compute the full-track signal-to-noise ratio (SNR) on each stem. The results are shown in \Cref{tab:results}. 
For the split setup, the MX-* result was evaluated using the sum of \mbox{MX-I} and MX-V estimates. Although the MX stem contents are different from the canonical DnR v3, the results indicated that the performance of the model is within a similar range to the benchmark reported in \cite{Watcharasupat2024RemasteringDivideRemaster}. The model interestingly did not exhibit any overfitting behavior during training, despite MUSDB18-HQ and MoisesDB combined being significantly smaller than the subset of FMA used in \cite{Watcharasupat2024RemasteringDivideRemaster}.

% \begin{table}[t]
%     \centering
%     \footnotesize
%     \setlength{\tabcolsep}{2.5pt}
%     \begin{tabularx}{\columnwidth}{X*{8}{S[table-format=2.2]}}
%     \toprule 
%     & \multicolumn{4}{c}{\textbf{SNR}} & \multicolumn{4}{c}{\textbf{SI-SNR}} \\
%     \cmidrule(lr){2-5}\cmidrule(lr){6-9}
%     \textbf{Model} & {DX} & {MX-I} & {MX-V} & {FX}
%     & {DX} & {MX-I} & {MX-V} & {FX}\\
%     \midrule
%     Bandit    & 14.3 & 10.1 & 9.4 & 10.1 & 14.1 & 9.7 & 8.9 & 9.7 \\
%     Banquet     & \bfseries 14.9 & \bfseries 10.6 &\bfseries 9.9 & \bfseries 10.4  & \bfseries 14.8 & \bfseries 10.2 &\bfseries 9.4 &\bfseries 10.0 \\
%     \midrule
%     Cohen's $d$ & 
%     0.72 & 0.38 & 0.71 & 0.21 & 
%     0.73 & 0.38 & 0.73 & 0.24\\
%     \bottomrule
%     \end{tabularx}
%     \caption{Median SNR and SI-SNR of each model.}
%     \label{tab:results}
% \end{table}

\begin{table}[t]
    \centering
    % \tiny
    \footnotesize
    \begin{tabularx}{\columnwidth}{lX*{5}{S[table-format=2.2]}}
    \toprule 
    \textbf{Setup} & \textbf{Model} &  \textbf{DX} & \textbf{MX-V} & \textbf{MX-I} & \textbf{MX-*} & \textbf{FX}\\
    \midrule
Inst. only & Bandit (37.0 M) & 16.0 &  & 11.6 &  & 11.2 \\
 & Banquet (19.7 M) & 16.1 &  & 11.8 &  & 11.3 \\
 & Cohen's \textit{d} & 0.16 &  & 0.11 &  & 0.14 \\
    \midrule
Combined & Bandit (37.0 M) & 14.7 &  &  & 11.4 & 10.2 \\
 & Banquet (19.7 M) & 14.9 &  &  & 11.6 & 10.3 \\
 & Cohen's \textit{d} & 0.34 &  &  & 0.23 & 0.08 \\
    \midrule
Split & Bandit (37.0 M) & 14.3 & 9.4 & 10.1 & 11.4 & 10.1 \\
 & Banquet (19.7 M) & 14.9 & 9.9 & 10.6 & 11.8 & 10.4 \\
 & Cohen's \textit{d} & 0.72 & 0.71 & 0.38 & 0.42 & 0.21 \\
    \bottomrule
    \end{tabularx}
    \caption{Median SNR of Bandit and Banquet in different training setups. Paired-sample Cohen's \textit{d} values indicate effect sizes of Banquet performance relative to Bandit.}
    \label{tab:results}
\end{table}

Across all setups and stems, the Banquet model performed statistically significantly better ($p < 0.01$) than Bandit, although with varying effect sizes. In the instrumental-only setup, the effect sizes were all very small, with the models performing within \SI{0.2}{dB} of each other, indicating that either model would work similarly in this setup. With the inclusion of vocals in the MX stem, the performance dropped by around \SI{1}{dB} for DX and FX, indicating that this setup is likely harder. Between the models, the performances are still with \SI{0.2}{dB} of each other, but the effect sizes are slightly higher in the DX and MX stems than in the instrumental MX setup while the effect size of the FX stem is very small. When treating the singing vocals and instrumentals separately, the Bandit model dropped in performance by \SI{0.4}{dB} on the speech stem. Across the board, moderate-to-large effect sizes were seen in DX and all MX stems, with the Banquet model performing \SI{0.4}{dB} to \SI{0.6} dB better. 

\begin{figure}
    \centering
    \includegraphics[width=0.5\columnwidth]{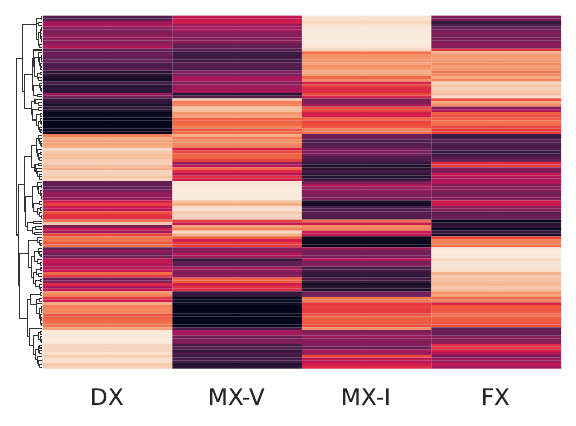}
    \vspace{-3ex}
    \caption{Normalized clustermap of Banquet's $\bm{\gamma}_i$ for each stem in the split MX setup. Lighter color indicates larger normalized value. }
    \label{fig:feature}
\end{figure}

Although both models consist of a single encoder responsible for computing the mixture embedding $\mathbf{\Upsilon}$, Bandit has a dedicated decoder for each stem while Banquet has a single shared decoder. As a result, in Bandit, the ``separation'' happens in the band-wise decoding block that maps $\mathbf{\Lambda}_b$ to $\mathbf{M}_{i, b}$, where $i$ is the stem index and $b$ is the band index. This map is nonlinear, likely allowing the representation $\mathbf{\Upsilon}$ to remain entangled. In Banquet, the ``separation'' occurs in the band-agnostic FiLM layer. Given that FiLM is equivalent to an affine operation whose linear map is constrained to a diagonal matrix, Banquet likely encourages independence across features or groups thereof. 
This conjecture is partially supported by the 
 \textit{z}-normalized cluster map of $\bm{\gamma}_i$, shown in \Cref{fig:feature}. Although the clustering is not fully obvious, it can be seen that most features are only activated for one or two of the stems, likely indicating that each abstract feature is responsible for a semantic concept specific to only one or two stems. 

\section{Conclusion}

In this work, we demonstrate a straightforward extension of the Bandit and Banquet models for a CASS setup that distinguishes between singing voice, dialogues, and instrumental music. Experimental results indicated that the query-based Banquet model performed significantly better despite only requiring half the parameters. Additional analyses are required to better understand these behaviors.

\newpage
\section{Acknowledgments}

This work was done while K.~N.~Watcharasupat was supported by the IEEE Signal Processing Society Scholarship Program.

The authors would like to thank Darius P{\'e}termann; 
Pablo Delgado and the Netflix Machine Learning Platform team; and, William Wolcott and the Netflix Audio Algorithms team for their assistance with the project.

\section{Ethics Statement}

The four-stem models developed in this work utilized training data derived from MUSDB18-HQ and MoisesDB, both permitting only non-commercial research use. Therefore, these models were developed strictly for research purposes only and will not be used in production at Netflix. Model weights for these systems will be released on a strictly non-commercial license. 

\small
% For bibtex users:
\bibliography{references}

\end{document}